\documentclass[structabstract]{aa}  
\usepackage{graphicx}
\usepackage{txfonts}
\begin{document}

  \titlerunning{Foreground neutral gas towards OMC-2}
   \title{\textit{Herschel} CHESS discovery of the fossil cloud that gave birth to the Trapezium and Orion~KL}


   \author{A. L\'opez-Sepulcre
          \inst{1}
          \and
          M. Kama\inst{2}
\and
          C. Ceccarelli\inst{1}
\and
          C. Dominik\inst{2,}\inst{3}
\and
          E. Caux\inst{4,}\inst{5}
\and
          A. Fuente\inst{6}
\and
          T. Alonso-Albi\inst{6}                    
          }

   \institute{UJF-Grenoble 1 / CNRS-INSU, Institut de Plan\'{e}tologie et d\textquoteright Astrophysique de Grenoble (IPAG) UMR 5274, Grenoble, F-38041, France
              \email{ana.sepulcre@obs.ujf-grenoble.fr}
         \and
             Astronomical Institute Anton Pannekoek, University of Amsterdam, Amsterdam, The Netherlands
         \and
             Department of Astrophysics/IMAPP, Radboud University Nijmegen, Nijmegen, The Netherlands             
         \and
             Universit\'e de Toulouse, UPS-OMP, IRAP, Toulouse, France
         \and
             CNRS, IRAP, 9 Av. colonel Roche, BP 44346, 31028 Toulouse Cedex 4, France    
         \and
             Observatorio Astron\'omico Nacional, P.O. Box 112, 28803 Alcal\'a de Henares, Madrid, Spain
             }

   \date{Received ; accepted }

 
  \abstract
   {The Orion~A molecular complex is a nearby (420~pc), very well studied stellar nursery that is believed to contain examples of triggered star formation.}
   {As part of the \textit{Herschel} Guaranteed Time Key Programme CHESS, we present the discovery of a diffuse gas component in the foreground of the intermediate-mass protostar OMC-2~FIR~4, located in the Orion~A region.}
   {Making use of the full HIFI spectrum of OMC-2~FIR~4 obtained in CHESS, we detected several ground-state lines from OH$^+$, H$_2$O$^+$, HF, and CH$^{+}$, all of them seen in absorption against the dust continuum emission of the protostar's envelope. We derived column densities for each species, as well as an upper limit to the column density of the undetected H$_3$O$^+$. In order to model and characterise the foreground cloud, we used the Meudon PDR code to run a homogeneous grid of models that spans a reasonable range of densities, visual extinctions, cosmic ray ionisation rates and far-ultraviolet (FUV) radiation fields, and studied the implications of adopting the Orion Nebula extinction properties instead of the standard interstellar medium ones.}
   {The detected absorption lines peak at a velocity of 9~km~s$^{-1}$, which is blue-shifted by 2~km~s$^{-1}$ with respect to the systemic velocity of OMC-2~FIR~4 ($V_\mathrm{LSR} = 11.4$~km~s$^{-1}$). The results of our modelling indicate that the foreground cloud is composed of predominantly neutral diffuse gas ($n_\mathrm{H} = 100$~cm$^{-3}$) and is heavily irradiated by an external source of FUV that most likely arises from the nearby Trapezium OB association. The cloud is 6 pc thick and bears many similarities with the so-called C$^{+}$ interface between Orion-KL and the Trapezium cluster, 2~pc south of OMC-2~FIR~4.}
   {We conclude that the foreground cloud we detected is an extension of the C$^{+}$ interface seen in the direction of Orion~KL, and interpret it to be the remains of the parental cloud of OMC-1, which extends from OMC-1 up to OMC-2.}

  
  \maketitle
%

\section{Introduction}\label{intro}

The Orion region contains the nearest giant molecular clouds to the Sun and has been the object of a great variety of studies in the past few decades. Within this region lies the Orion~A molecular cloud, a very active site of star formation, at a distance of $\sim$420~pc (Hirota et al.~\cite{hirota07}, Menten et al.~\cite{menten07}). Located in the interior of a 300~pc diameter superbubble of gas inflated over the past 10~Myr by various OB associations (Reynolds \& Ogden \cite{rey79}), Orion~A is exposed to external forces that are believed to compress it and trigger star formation (e.g. Shimajiri et al. \cite{shim11}). 

The northern portion of the Orion~A complex is associated with the emblematic Orion Nebula (M42) and the Trapezium OB stellar cluster. Immediately behind the nebula lies the Orion Molecular Cloud 1 (OMC-1), which contains the BN infrared (IR) cluster and the KL molecular clump (see also the sketch in Fig.~\ref{fgeo}b). An integral-shaped filament (ISF) of dense molecular gas extends towards the north and south of OMC-1, with a length of about 15~pc and a relatively small width of $\sim$0.5~pc (Bally et al. \cite{bally87}, Johnstone \& Bally \cite{john99}). A velocity gradient is observed along the filament starting from $V_\mathrm{LSR} = 4$~km~s$^{-1}$ in the southernmost edge, up to $V_\mathrm{LSR} = 12$~km~s$^{-1}$ in the northern part, with an abrupt change in M42. This velocity gradient is interpreted by Kutner et al. (\cite{kut77}) as rotation of the entire cloud about an east-west axis located roughly at the position of OMC-1. The northern region of the ISF comprises two main molecular clouds active in star formation, known as OMC-2 and OMC-3. They are bound to the north by the reflection nebula NGC~1977, and have been extensively investigated at several wavelengths (see Peterson \& Megeath \cite{pet08} for a review).

Here we report the discovery of a diffuse gas component in the foreground of OMC-2. The data are part of the \textit{Herschel}\footnote{\textit{Herschel} is a European Space Agency (ESA) space observatory with science instruments provided by European-led principal investigator consortia and with important participation from the National Aeronautics and Space Administration (NASA).} Guaranteed Time Key Programme Chemical HErschel Surveys of Star forming regions (CHESS\footnote{http://www-laog.obs.ujf-grenoble.fr/heberges/chess/};  Ceccarelli et al.~\cite{cec10}), which has conducted unbiased spectral surveys of eight star-forming regions, each representative of a particular aspect of the star formation process: evolutionary stage, mass of the forming star, and/or interaction with the surroundings. Among the CHESS targets there is OMC-2~FIR~4, an intermediate-mass protostar and the brightest sub-millimetre source in OMC-2 (Johnstone \& Bally \cite{john99}, Kama et al. \textit{in prep.}). The rich spectral dataset provided by the HIFI spectrograph has allowed us to detect several ground-state lines from OH$^{+}$, H$_2$O$^{+}$, HF, and CH$^+$, seen in absorption against the dust continuum emission of the protostars's envelope.

We describe these observations in Sect.~\ref{obs}. The spectra and derived physical quantities are presented in Sect.~\ref{results}. Section~\ref{model} describes the modelling of the observations. In Sect.~\ref{discussion} we discuss the possible interpretations of our results, and a final summary is presented in Sect.~\ref{conclusion}.


\section{Observations}\label{obs}

The observations towards OMC-2~FIR~4 ($\alpha_\mathrm{2000} = 05^\mathrm{h}35^\mathrm{m}26.97^\mathrm{s}$, $\delta_\mathrm{2000} = -05^\mathrm{d}09^\mathrm{m}54.5^\mathrm{s}$) were carried out with the HIFI spectrograph on-board the \textit{Herschel} satellite (de Graauw et al. \cite{graauw10}, Pilbratt et al.~\cite{pil10}) in single-pointing, dual-beam switch mode and with a redundancy of 8. The wide band spectrometer (WBS), whose spectral resolution is 1.1~MHz, was used. The double sideband (DSB) spectra were reduced and deconvolved using  HIPE\footnote{HIPE is a joint development by the Herschel Science Ground Segment Consortium, which consists of the ESA, the NASA Herschel Science Center, and HIFI, the Photodetector Array Camera and Spectrometer, and the Spectral and Photometric Imaging Receiver consortia.} version 7.0, and were subsequently exported to GILDAS\footnote{http://www.iram.fr/IRAMFR/GILDAS} for analysis. The final spectra presented here are equally weighted averages of the H and V polarisations, smoothed to a velocity channel spacing of 0.5~km~s$^{-1}$.

\begin{table*}[!hbt]
\caption{Parameters of the observed absorption lines.}
\begin{tabular}{lcccccccccc}
\hline
Line & $\nu$ & $A_\mathrm{ul}$ & $\eta_\mathrm{beam}$ & $\theta_\mathrm{beam}$ & 1$\sigma$ RMS & $T_\mathrm{cont}$ & $V_\mathrm{LSR}$$^\mathrm{a}$ & $FWHM$$^\mathrm{a}$ & $\int \tau \mathrm{d}V_\mathrm{LSR}$ & $N_\mathrm{mol}$\\
 & (MHz) & ($10^{-2}$~s$^{-1}$) & (\%) & ($''$) & (mK) & (K) & (km~s$^{-1}$) & (km~s$^{-1}$) & (km~s$^{-1}$) & ($10^{12}$ cm$^{-2}$)\\
\hline
\multicolumn{11}{c}{OMC-2~FIR~4}\\
\hline
CH$^{+}$(1--0) & 835137.5 & 0.64 & 74.7 & 26 & 41 & 0.48 & $9.5 \pm 0.1$ & $6.1 \pm 0.2$ & $12.0 \pm 1.6$ & $34.4 \pm 4.5$\\
OH$^{+}$(1$_{0,1/2}$--0$_{1,3/2}$) & 909158.8 & 1.05 & 74.4 & 24 & 33 & 0.54 & $9.3 \pm 0.1$ & $2.6 \pm 0.4$ & $1.1 \pm 0.4$ & $23.1 \pm 7.4$\\
OH$^{+}$(1$_{2,5/2}$--0$_{1,3/2}$)$^\mathrm{b}$ & 971803.8 & 1.82 & 74.1 & 22 & 95 & 0.58 & $9.2 \pm 0.1$ & $3.3 \pm 0.3$ & $5.0 \pm 1.6$ & $23.8 \pm 7.3$\\
OH$^{+}$(1$_{1,1/2}$--0$_{1,1/2}$)$^\mathrm{b}$ & 1032997.9 & 1.41 & 74.1 & 24 & 52 & 0.63 & $9.4 \pm 0.1$ & $2.9 \pm 0.4$ & $1.1 \pm 0.4$ & $24.9 \pm 7.9$\\
OH$^{+}$(1$_{1,3/2}$--0$_{1,3/2}$)$^\mathrm{b}$ & 1033118.6 & 1.76 & 74.1 & 24 & 52 & 0.63 & $9.3 \pm 0.2$ & $3.2 \pm 0.5$ & $2.6 \pm 0.6$ & $22.7 \pm 4.8$\\
\textit{o}-H$_2$O$^{+}$(1$_{1,1}$--0$_{0,0}$)$^\mathrm{d}$ & 1115204.0 & 3.02 & 63.7 & 19 & 72 & 0.68 & $8.2 \pm 0.3$ & $4.4 \pm 0.8$ & $1.8 \pm 0.4$ & $7.6 \pm 1.9$\\
H$_3$O$^{+}$(0$_{01}$--1$_{00}$)$^\mathrm{c}$ & 984708.7 & 2.30 & 74.1 & 22 & 52 & 0.59 & --- & --- & $< 0.8$ & $< 0.8$\\
HF(1--0) & 1232476.3 & 2.42 & 63.7 & 17 & 130 & 0.75 & $10.0 \pm 0.1$ & $2.8 \pm 0.3$ & $4.8 \pm 1.4$ & $12.3 \pm 3.4$\\
\hline
\multicolumn{11}{c}{OMC-3~MM6}\\
\hline
OH$^{+}$(1$_{0,1/2}$--0$_{1,3/2}$)$^\mathrm{b}$ & 909158.8 & 1.05 & 74.4 & 24 & 13 & 0.58 & --- & --- & $< 0.1$ & $< 3.0$\\
\hline
\end{tabular}
\begin{itemize}
\item[$^\mathrm{a}$]{Measured from a Gaussian fit to the line}
\item[$^\mathrm{b}$]{Blended with another hyperfine component (see Fig.~\ref{fspt}); $N_\mathrm{mol}$ estimated assuming that all flux comes from this component}
\item[$^\mathrm{c}$]{Undetected: 3$\sigma$ upper limits are given for $\int \tau \mathrm{d}V_\mathrm{LSR}$ and $N_\mathrm{mol}$, assuming $FWHM = 2.6$~km~s$^{-1}$}
\item[$^\mathrm{d}$]{$F = 5/2 - 3/2$ transition; rest frequency from M\"urtz et al.~\cite{murtz98}}
\end{itemize}
\label{tobs}
\end{table*}

Table~\ref{tobs} lists the observational parameters for each absorption line detected towards OMC-2~FIR~4: beam efficiency, beam width, and the measured 1$\sigma$ RMS in main beam temperature, $T_\mathrm{mb}$.

Following the detection towards OMC-2~FIR~4, as part of a \textit{Herschel} OT2 accepted project, we recently observed OH$^{+}$(1$_{0,1/2}$--0$_{1,3/2}$) towards OMC-3~MM6, located $\sim$2~pc farther north, to evaluate the extent of the absorbing foreground layer. The  observations were made on 22 February 2012 in DSB, single-pointing mode, and were subsequently analysed with HIPE version 8.1. The corresponding observational parameters are also listed in Table~\ref{tobs}. Even though the RMS level is lower than for OMC-2~FIR~4 and the continuum level is slightly higher, the line is not detected towards this source.

\section{Results} \label{results}

\subsection{Spectra}

\begin{figure*}[!th]
\begin{center}
\begin{tabular}{lr}
\includegraphics[angle=0,scale=.7]{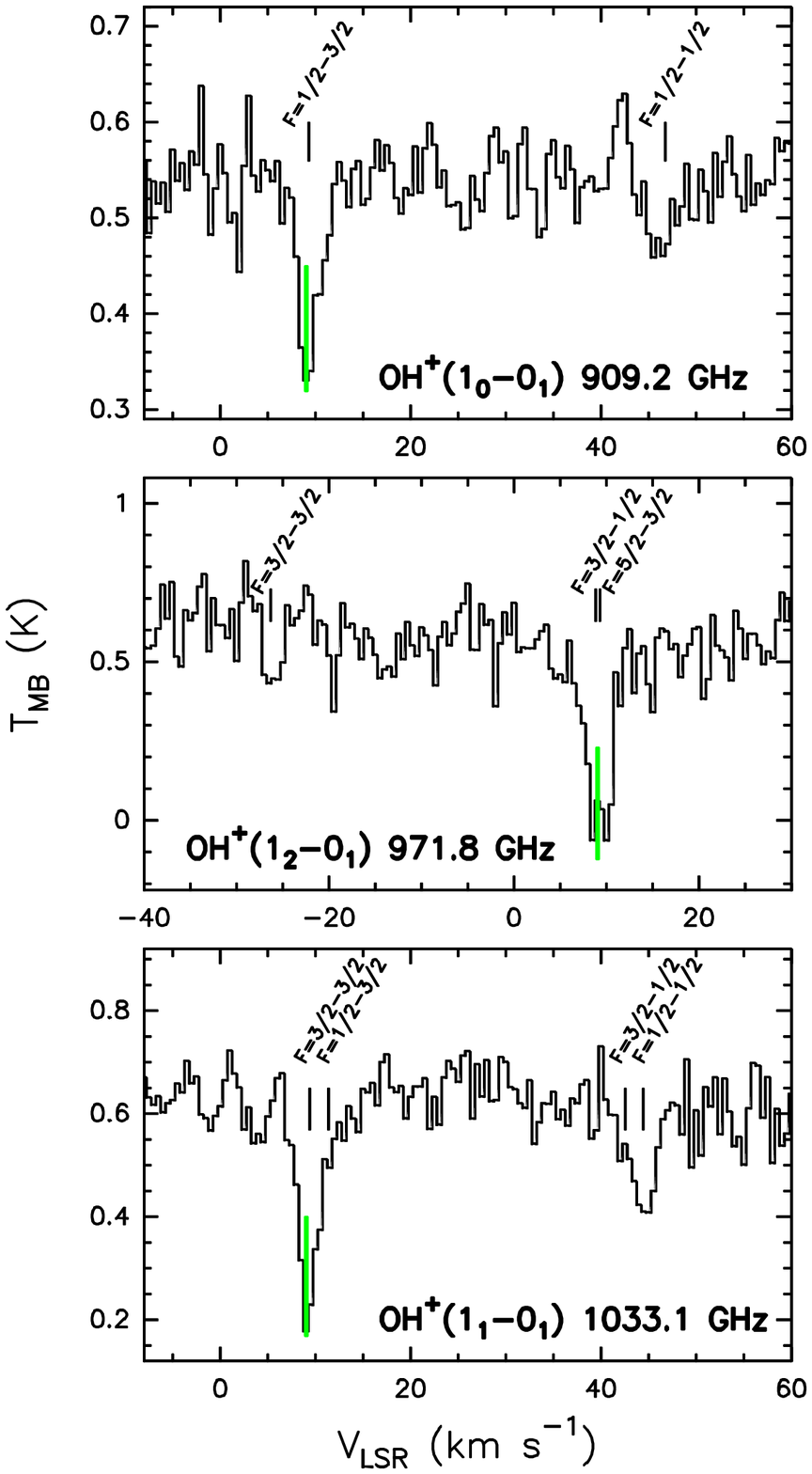} & \includegraphics[angle=0,scale=.7]{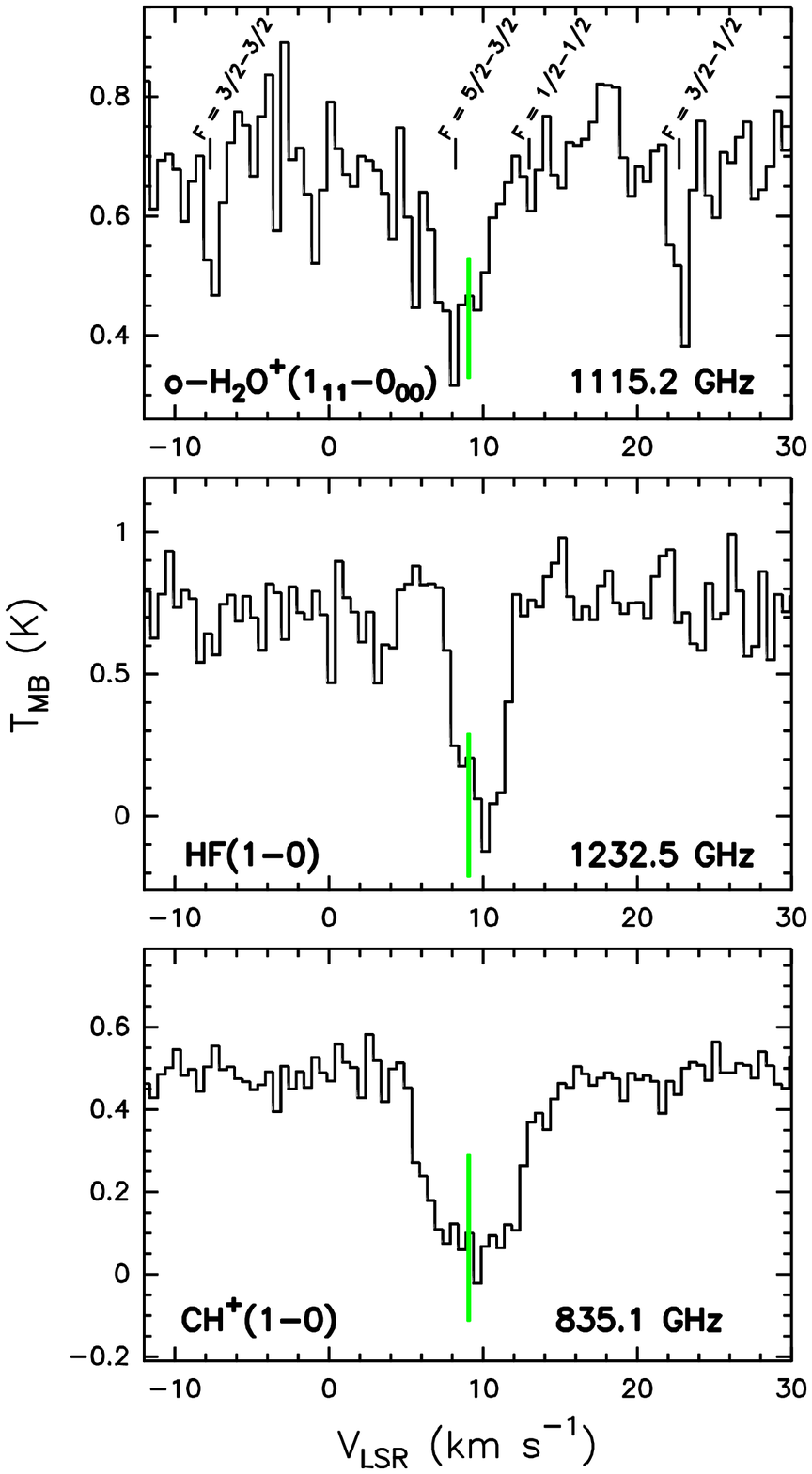}\\
\end{tabular}
\caption{Spectral lines detected in absorption towards OMC-2~FIR~4. The green vertical lines mark $V_\mathrm{LSR} = 9.3$~km~s$^{-1}$, i.e. the systemic velocity of the OH$^{+}$ lines.}
\label{fspt}
\end{center}
\end{figure*}

Figure~\ref{fspt} shows the observed absorption lines towards OMC-2~FIR~4, all of them corresponding to ground-state transitions. Four OH$^{+}$ lines, three of which are blends of two different hyperfine components, are detected with a signal-to-noise ratio greater than 3. On the other hand, \textit{o}-H$_2$O$^{+}$(1$_{1,1,5/2}$--0$_{0,0,3/2}$) is detected at 4$\sigma$, and H$_3$O$^{+}$, whose ground-state transition falls at 984.7~GHz, is undetected. Both HF(1--0) and CH$^+$(1--0) display clear absorption and reach the saturation level.

Two features can be noticed from Fig.~\ref{fspt} (see also Table~\ref{tobs}). Firstly, the absorption lines peak around a velocity of 9~km~s$^{-1}$, which is shifted by more than 2~km~s$^{-1}$ from that of OMC-2~FIR~4 ($V_\mathrm{LSR} = 11.4$~km~s$^{-1}$). Secondly, similar studies recently performed towards other sources typically reveal absorption over a wide range of velocities, often several tens of km~s$^{-1}$ (e.g. Gerin et al. \cite{gerin10}, Neufeld et al. \cite{neu10_O}, Monje et al. \cite{monje11}), and in some cases they are even contaminated by blending emission lines (e.g. Phillips et al. \cite{phil10}). This is not the case here, where the lines have relatively small measured line widths and apparently trace one single velocity component.

The centre of the \textit{o}-H$_2$O$^+$(1$_{1,1,5/2}$--0$_{0,0,3/2}$) line is blue-shifted with respect to that of the OH$^+$ lines by 1~km~s$^{-1}$ (see Table~\ref{tobs}). This is most likely due to the uncertainties in the rest frequency measurements reported in the literature for H$_2$O$^+$ (see, e.g. Table~1 in Ossenkopf et al.~\cite{ossen10}). Here we have taken the rest frequencies from M\"urtz et al.~\cite{murtz98}.

As mentioned above, OH$^+$ is not detected towards OMC-3~MM6 down to the noise level of the observations.

\subsection{Column density estimates}\label{coldens}

The molecular column densities of each species displaying absorption lines were estimated assuming that the particles are in the ground state, since their collisional coefficients are unknown. Under this assumption, the column density, $N_\mathrm{mol}$, is given by

\begin{equation}
N_\mathrm{mol} = 8 \pi \frac{1}{A_\mathrm{ul}} \frac{g_\mathrm{l}}{g_\mathrm{u}}
\frac{\nu^3}{c^3} \int \tau \mathrm{d}V_\mathrm{LSR}~,
\label{eN}
\end{equation}
where $A_\mathrm{ul}$ is the spontaneous emission Einstein coefficient,
$g_\mathrm{l}$ and $g_\mathrm{u}$ are the statistical weights of the
lower and upper energy levels, respectively, and $c$ is the speed of light.

Since the ground state 
of both OH$^+$ and \textit{o}-H$_2$O$^+$ are split into two levels with a negligible 
energy difference, we assume that the population of each is proportional to their 
corresponding level degeneracies, namely $g_\mathrm{l} = 4$ for $F = 3/2$ and 
$g_\mathrm{l} = 2$ for $F = 1/2$. We therefore included the corresponding factor in Eq.~\ref{eN} 
to obtain the total column density of each of these two species.

\begin{figure}[!hbt]
\begin{center}
\includegraphics[angle=0,scale=.45]{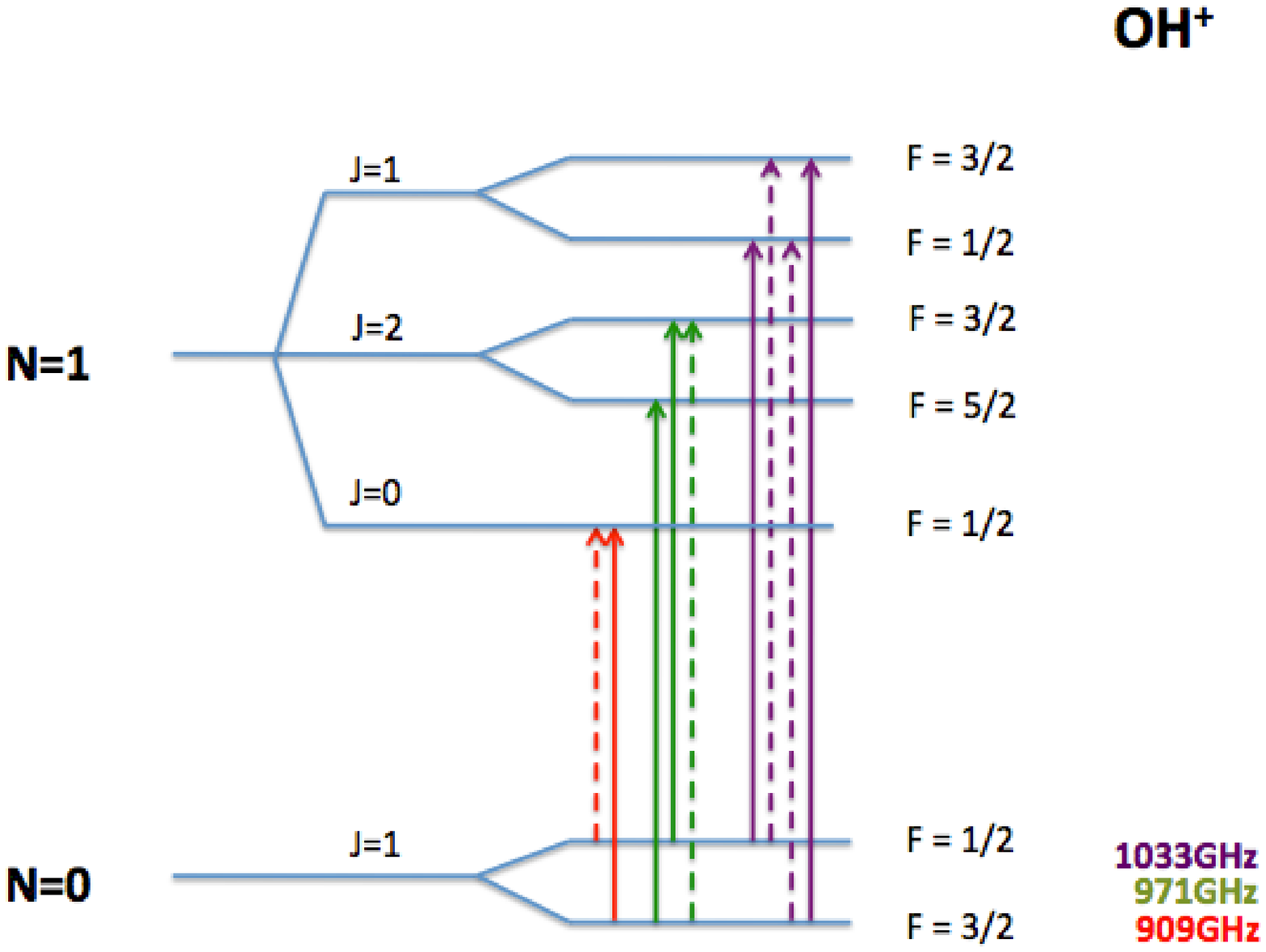}
\caption{OH$^{+}$ energy level digram for the detected transitions. Solid arrows depict lines we observe at an S/N~$> 3$, while dashed arrows mark the lines with an S/N~$< 3$.}
\label{fohp}
\end{center}
\end{figure}

A sketch of the energy levels involved in the OH$^{+}$ 
$N = 0 \rightarrow 1$ transition is shown in Fig.~\ref{fohp}.

The optical depth at a given velocity channel, $\tau$, is given by

\begin{equation}
\tau = -\ln \left(\frac{T_\mathrm{line}}{T_\mathrm{cont}}\right)~.
\label{etauabs}
\end{equation} 

The resulting velocity-integrated optical depths and column densities are listed in Table~\ref{tobs}, together with some parameters of the transitions. The table includes upper limits for the undetected H$_3$O$^+$ transition. The errors of the derived column densities are of the order of 30\% and include the contribution of the spectral noise around the lines as well as a 25\% uncertainty on the data calibration (Roelfsema et al. \cite{roe12}).

In the case of OH$^{+}$, we used the four lines that are clearly detected above 3$\sigma$, all of which yield very similar values. The final OH$^{+}$ column density we adopted is the arithmetic average of the four column-density derivations, which amounts to $N_\mathrm{OH^+} = (2.4 \pm 0.4) \times 10^{13}$~cm$^{-2}$. We computed 3$\sigma$ upper limits to the column densities of their corresponding species for the undetected OH$^+$ and H$_2$O$^+$ hyperfine components. They all exceed the values listed in Table~\ref{tobs}, and are therefore consistent with them.

The highest column density of the three O-bearing species corresponds to OH$^{+}$, followed by H$_2$O$^{+}$, and finally H$_3$O$^{+}$, with $N_\mathrm{OH^{+}}/N_\mathrm{H_2O^+} = 3.1$, and $N_\mathrm{H_2O^+}/N_\mathrm{H_3O^+} > 9.5$.

\section{Modelling}\label{model}

To model the chemistry of OH$^{+}$, H$_2$O$^{+}$, HF, CH$^+$, and the undetected H$_3$O$^{+}$, and to determine the physical
properties that best match our observations, we made use of the
Meudon photo-dissociation (PDR) code\footnote{http://pdr.obspm.fr/PDRcode.html}, developed by Le Petit et
al.~(\cite{petit06}).  The model computes
the chemical and thermal structure of a homogenous plane-parallel
slab of gas and dust illuminated by a radiation field coming from one
or both sides of the cloud.  We adopted the standard carbon and
oxygen chemical network, plus the network for the fluorine chemistry
following Neufeld et al. (\cite{neu09}). The adopted elemental abundances are $1.3 \times 10^{-4}$ (carbon), $3.2 \times
10^{-4}$ (oxygen) and $1.8 \times 10^{-8}$ (fluorine).  We considered a slab
illuminated by the interstellar standard radiation field (ISRF;
1.7~G$_0$, where G$_0 = 5.9 \times 10^{-14}$~erg~cm$^{-3}$ is Habing's
ISRF) on one side --since we do not expect multiple UV sources to lie
close-by, which might irradiate heavily on both sides-- and by a stronger
field on the other side.

As can be seen in Fig.~\ref{fspt}, the CH$^{+}$(1--0) and HF(1--0) lines are 
saturated and have larger widths, suggesting the presence of more than one 
kinematical component. For this reason we based the comparison between models 
and observations (see below) on the lines produced by OH$^{+}$ and 
H$_2$O$^{+}$, which are optically thinner and narrower and appear to trace a 
single velocity component. The upper limit on the H$_3$O$^{+}$ column density 
also places a strong constraint on the best-fitting model.

Our grid of models spans a sufficiently large parameter space to find the solution that best
fits the observations, as described below.

\subsection{Grid of models}\label{grid}

We ran two grids of models, a first one (coarse grid) sampling a large parameter
space in gas density ($n_\mathrm{H}$), visual extinction
($A_\mathrm{V}$), incident far-UV (FUV) radiation field on one side of
the slab ($u_\mathrm{FUV}$), and cosmic ray ionisation rate
($\zeta_\mathrm{cr}$), and then a finer grid around the best fit from the coarse grid. 

We probed FUV field fluxes up to 1700~G$_0$ 
because strong radiation fields are expected in this region and are also consistent 
with previous work, which claims that large amounts of UV radiation
are necessary to explain CO and C$^+$ observations towards the OMC-2
region (Herrmann et al. \cite{her97}, Crimier et
al. \cite{crim09}). Similarly, we covered a large range of $\zeta_\mathrm{cr}$ 
values, as previous work suggested enhancement with respect to the canonical value 
(Gupta et al. \cite{gupta10}). The range of gas densities chosen covers typical
diffuse interstellar cloud values, and we also selected
relatively low values of $A_\mathrm{V}$ because the slab does not
display a distinct CO line component (Shimajiri et al. \cite{shim11}).
The range of parameters of the two grids are listed in
Table~\ref{tmodel}, together with other parameters, such as elemental abundances 
and grain properties, which we kept fixed for all models.

\begin{table}[!h]
\caption{Parameters of the Meudon PDR models}
\begin{tabular}{lll}
\hline
 & First grid & Second grid\\
 \hline
$n_\mathrm{H}$ (cm$^{-3}$) & 10, 100, 1000 & 100, 200, 300\\
$A_\mathrm{V}$ (mag) & 0.5, 1, 3 & 1, 1.5\\
$u_\mathrm{FUV}$ (G$_0$) & 1.7, 17, 170, 1700 & 170, 340, 510, 850, 1190, 1700\\
$\zeta_\mathrm{cr}$ ($10^{-17}$~s$^{-1}$) & 1, 10, 100,1000 & 10, 30\\
\hline
 & \multicolumn{2}{l}{Elemental abundances wrt H nuclei}\\ 
\hline
C & \multicolumn{2}{l}{1.32$\times 10^{-4}$}\\
O & \multicolumn{2}{l}{3.19$\times 10^{-4}$}\\
N & \multicolumn{2}{l}{7.50$\times 10^{-5}$}\\
F & \multicolumn{2}{l}{1.80$\times 10^{-8}$}\\
S & \multicolumn{2}{l}{1.86$\times 10^{-5}$}\\
\hline
 & \multicolumn{2}{l}{Other fixed parameters}\\
\hline
$R_\mathrm{V}$ & \multicolumn{2}{l}{3.1}\\
$N_\mathrm{H}/A_\mathrm{V}$ & \multicolumn{2}{l}{$1.9 \times 10^{21}$ ~mag~cm$^{-2}$}\\
Initial $T_\mathrm{gas}$ & \multicolumn{2}{l}{70~K}\\
Grain radii & \multicolumn{2}{l}{$3 \times 10^{-7}$ - $3 \times 10^{-5}$~cm}\\
\hline
\end{tabular}
\label{tmodel}
\end{table}

Cardelli et al.~(\cite{cardelli89}) reported $R_\mathrm{V} = 5.5$ and $N_\mathrm{H}/A_\mathrm{V} = 2.9\times 10^{21}$~mag~cm$^{-2}$ towards the Trapezium stars within the Orion Nebula. Since we are not certain that these values can be extended to the OMC-2 cloud, we adopted standard interstellar medium values (see Table~\ref{tmodel}). However, to see the effect of changing these two parameters on the results, we ran a number of test models around the best fitting one with the Trapezium $R_\mathrm{V}$ and $N_\mathrm{H}/A_\mathrm{V}$ values. The differences are discussed in Sect.~\ref{reliability}.

\subsection{Best-fit model}\label{fit}

Considering the coarse grid of models, only two solutions are found to lie close to the observed column densities of OH$^{+}$,
H$_2$O$^{+}$ and H$_3$O$^{+}$:

\begin{itemize}
\item[$\bullet$] $A_\mathrm{V}=1$~mag, $n_\mathrm{H}=10$~cm$^{-3}$, $\zeta_\mathrm{cr}=10^{-17}$~s$^{-1}$, $u_\mathrm{FUV}=170$~G$_0$
\item[$\bullet$] $A_\mathrm{V}=1$~mag, $n_\mathrm{H}=100$~cm$^{-3}$, $\zeta_\mathrm{cr}=10^{-16}$~s$^{-1}$, $u_\mathrm{FUV}=1700$~G$_0$~.
\end{itemize}

The latter is much more likely, since the former implies having a
60-parsec-deep slab of material between OMC-2~FIR~4 and us but with a
remarkably low velocity dispersion. We therefore discarded this and consider only the $n = 100$~cm$^{-3}$ solution below. A posteriori 
we also show that the the high FUV field 
irradiation implied by this solution is also consistent (Sect.~\ref{geo}).

To better constrain the parameters of the absorbing slab, we
then ran a second grid of models that covered a finer-tuned set of values
around those of the best solution, in total 72 additional models
(Table~\ref{tmodel}, right column). The models with $A_\mathrm{V} = 1.5$~mag and/or 
$n_\mathrm{H} > 100$~cm$^{-3}$ predict a higher amount of
H$_3$O$^{+}$ than the upper limit we measured. On the other hand, the $A_\mathrm{V} = 0.5$~mag models 
from the first coarse grid underpredict the observed amount of OH$^{+}$. This leads us to
conclude that the visual extinction, $A_\mathrm{V}$, of the foreground
slab lies just around 1~magnitude and is not denser than 100~cm$^{-3}$.

The $\zeta_\mathrm{cr}=10^{-16}$~s$^{-1}$ models predict lower column densities than those observed, whereas the $\zeta_\mathrm{cr}=3 \times 10^{-16}$~s$^{-1}$ models match them within the errors.

\begin{figure}[!htb]
\begin{center}
\includegraphics[angle=0,scale=.27]{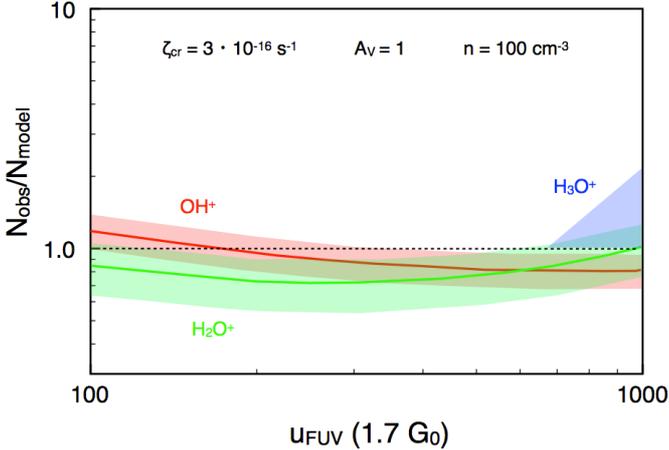}\\ 
\caption{Ratio of observed to modelled column densities, $N_\mathrm{obs}/N_\mathrm{model}$, of OH$^{+}$ (red), H$_2$O$^+$ (green) and H$_3$O$^+$ (blue), as a function of the FUV field for a slab with $A_\mathrm{V} = 1$, $n_\mathrm{H} = 100$~cm$^{-3}$, and $\zeta_\mathrm{cr} = 3 \times 10^{-16}$~s$^{-1}$. The widths of the red and green bands represent the errors around the measured column densities of OH$^{+}$ and H$_2$O$^+$. The blue area marks the only range of FUV field values possible for the models to be compatible with the derived 3$\sigma$ upper limit to the H$_3$O$^+$ column density. A horizontal black dotted line is depicted at $N_\mathrm{obs}/N_\mathrm{model} = 1$.}
\label{fcon2}
\end{center}
\end{figure}

In Fig.~\ref{fcon2} we present a plot of the ratio of observed to
modelled column densities, $N_\mathrm{obs}/N_\mathrm{model}$,
as a function of $u_\mathrm{FUV}$ for a 100~cm$^{-3}$ dense slab for which  
$\zeta_\mathrm{cr} = 3 \times 10^{-16}$~s$^{-1}$. 
While OH$^{+}$ and H$_2$O$^{+}$ converge around $N_\mathrm{obs}/N_\mathrm{model} = 1$ for the whole range of $u_\mathrm{FUV}$ plotted, the models overpredict the amount of H$_3$O$^+$ for values of $u_\mathrm{FUV}$ below $\sim 1200$~G$_0$. We are therefore 
looking at a diffuse cloud of gas that is being heavily irradiated by an external 
FUV field.

Table~\ref{tcolmod} lists the column densities predicted by the best-fit model ($u_\mathrm{FUV}=1700$~G$_0$) 
for the observed species and for other species of potential interest. 
The CH$^+$ column density
predicted by the best-fit model is lower than the observed value by a factor
$\sim$250, a remarkably high discrepancy. Indeed, none of
the models we ran (Table~\ref{tmodel}) reaches as close as a factor 20
below the observed column density of CH$^{+}$. This molecule is 
discussed in more detail in Sect.~\ref{chemistry}.

\begin{table}[!h]
\caption{Column densities predicted by the best-fit model with $u_\mathrm{FUV}=1700$~G$_0$ (see text)}
\begin{tabular}{lccc}
\hline
Species & \multicolumn{3}{c}{$N_\mathrm{mol}$ (cm$^{-2}$)}\\
\cline{2-4}
 & observed & modelled (ISM) & modelled (Orion)\\
 \hline
H$_2$ & --- & $6.1 \times 10^{19}$ & $3.1 \times 10^{19}$\\
H$_3$$^+$ & --- & $1.5 \times 10^{12}$ & $2.4 \times 10^{12}$\\
OH$^+$ & $(2.4 \pm 0.4) \times 10^{13}$ & $2.9 \times 10^{13}$ & $6.1 \times 10^{13}$\\
H$_2$O$^+$ & $(7.6 \pm 1.9) \times 10^{12}$ & $7.5 \times 10^{12}$ & $7.2 \times 10^{12}$\\
H$_3$O$^+$ & $< 8.0 \times 10^{11}$ & $3.5 \times 10^{11}$ & $1.2 \times 10^{11}$\\
C$^+$ & --- & $2.5 \times 10^{17}$ & $3.8 \times 10^{17}$\\
C & --- & $1.0 \times 10^{13}$ & $7.0 \times 10^{12}$\\
CO$^+$ & --- & $7.5 \times 10^{7}$ & $6.1 \times 10^{7}$\\
CO & --- & $3.1 \times 10^{9}$ & $7.4 \times 10^{8}$\\
HCO$^+$ & --- & $1.4 \times 10^{8}$ & $3.7 \times 10^{7}$\\
CH & --- & $4.7 \times 10^{8}$ & $2.6 \times 10^{8}$\\
CH$^+$ & $(3.4 \pm 4.5) \times 10^{13}$ & $7.7 \times 10^{10}$ & $1.3 \times 10^{11}$\\
C$_2$H & --- & $3.0 \times 10^{3}$ & $2.8 \times 10^{2}$\\
N$_2$H$^+$ & --- & $1.1 \times 10^{5}$ & $1.3 \times 10^{5}$\\
CN & --- & $2.4 \times 10^{6}$ & $1.2 \times 10^{6}$\\
HCN & --- & $2.7 \times 10^{3}$ & $1.3 \times 10^{3}$\\
HNC & --- & $4.0 \times 10^{2}$ & $6.6 \times 10^{1}$\\
HF & $(1.2 \pm 3.4) \times 10^{13}$ & $7.1 \times 10^{11}$ & $2.9 \times 10^{11}$\\ 
\hline
\end{tabular}
\label{tcolmod}
\end{table}

\subsection{Two HF velocity components}\label{hf}

What appears to be the
best fit for the O-bearing species fails to reproduce the amount of HF
observed, which is $\sim$20 times higher than the model
prediction. However, a closer look at the HF(1--0) spectrum 
in Fig.~\ref{fspt} suggests the presence of two different velocity components: 
a weak component around $V_\mathrm{LSR}$ 
of the foreground slab, 
and a stronger component at 10.5~km~s$^{-1}$, much closer to the 
systemic velocity of OMC-2~FIR~4.

According to our modelling, the amount of HF in the slab is around $10^{12}$~cm$^{-2}$.
The remaining $10^{13}$~cm$^{-2}$ (see Table~\ref{tobs}) presumably come from the denser OMC-2~FIR~4 boundary.
Indeed, we ran a few test PDR models for higher density gas ($n_\mathrm{H} = 10^4$~cm$^{-3}$) 
and high FUV irradiation fields (between 170 and 1700~G$_0$) and found column densities 
of $2 \times 10^{13}$~cm$^{-2}$ for HF. This is slightly higher than the value we measured, which may 
be explained by the fact that HF is depleted onto dust grains in denser molecular gas, as found by, 
e.g., Phillips et al. (\cite{phil10}). As for the O-bearing species, our test models predict low column densities, of $\sim 10^{11}$~cm$^{-2}$, 
consistent with having most of the flux of the observed lines coming from the slab and not from this
red-shifted component.

To summarise, our observations are compatible with the presence of two components: the foreground 
diffuse gas slab at a velocity of $\sim$9~km~s$^{-1}$, responsible for most of the OH$^+$ 
and H$_2$O$^{+}$ absorption, and the denser boundary of the OMC-2~FIR~4 core, where most
 of the observed HF is produced.
 
 \subsection{Reliability of the modelling}\label{reliability}

The reliability of the best-fit model is subject to the uncertainty in the values of some input parameters. In particular, as mentioned in Sect.~\ref{grid}, we investigated how the extinction properties affect the predicted column densities by running a number of test models around the best-fit values.

The last column in Table~\ref{tcolmod} lists the column densities of several species according to our best-fit model when adopting the values of $R_\mathrm{V}$ and $N_\mathrm{H}/A_\mathrm{V}$ reported in Cardelli et al.~(\cite{cardelli89}). These imply a deeper penetration of external dissociating and ionising photons and therefore a smaller amount of H$_2$ in the cloud, which alters the formation of other molecules. For the species considered in this study, this translates into an amount of HF lower by a factor 2.4, a higher amount of CH$^+$ by a factor 1.7, and a larger difference in the relative column densities of OH$^+$, H$_2$O$^+$ and H$_3$O$^+$, which causes OH$^+$ to be twice as abundant and H$_3$O$^+$ three times less abundant. The column density of H$_2$O$^+$ does not change significantly.

In summary, the use of the Orion Nebula extinction properties instead of the standard interstellar medium properties results in variations of at most a factor 3 in the predicted column densities. We find that, in order to better reproduce the observed column densities with the $R_\mathrm{V}$ and $N_\mathrm{H}/A_\mathrm{V}$ values from Cardelli et al.~(\cite{cardelli89}), it suffices to adopt a slightly lower cosmic ray ionisation rate of about $2 \times 10^{-16}$~s$^{-1}$. The low density and visual extinction, as well as the strong external FUV radiation field, need to be around the best-fit model values quoted in Sect.~\ref{fit}.
 
\section{Discussion}\label{discussion}

\begin{figure}[!h]
\begin{center}
\includegraphics[angle=0,scale=.5]{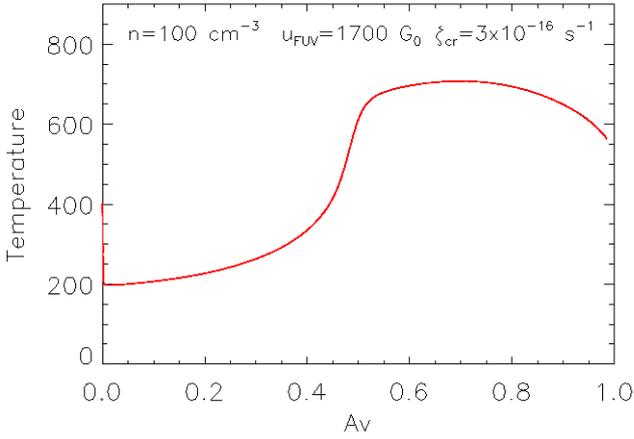} 
\caption{Gas temperature as a function of visual extinction, $A_\mathrm{V}$, predicted by the Meudon PDR code for a slab of gas with $n_\mathrm{H} = 100$~cm$^{-3}$, $\zeta_\mathrm{cr} = 3 \times 10^{-16}$~s$^{-1}$ and irradiated from the right by a FUV field of 1700~G$_0$.}
\label{ftemp}
\end{center}
\end{figure}

The results of our observations and modelling indicate that the OH$^{+}$ and H$_2$O$^{+}$ lines, as well as part of the HF, are tracing a foreground cloud of diffuse gas ($A_\mathrm{V} = 1$~mag, $n_\mathrm{H} = 100$~cm$^{-3}$), exposed to a cosmic ray ionisation rate $\zeta_\mathrm{cr} = 3 \times 10^{-16}$~s$^{-1}$ and a strong external FUV field of 1200-1700~G$_0$.  This causes the heavily irradiated side of the slab to have a considerably higher gas temperature than the other side (Fig.~\ref{ftemp}). With this heavy irradiation, UV pumping might have an effect on the level populations of the species under study, but given the lack of information about their collisional coefficients, we are not able to quantify it.

Below we discuss the chemistry and geometry of the cloud and propose an interpretation to explain its nature.

\subsection{Chemistry of the foreground slab}\label{chemistry}

\begin{figure}[!hbt]
\begin{center}
\includegraphics[angle=0,scale=.33]{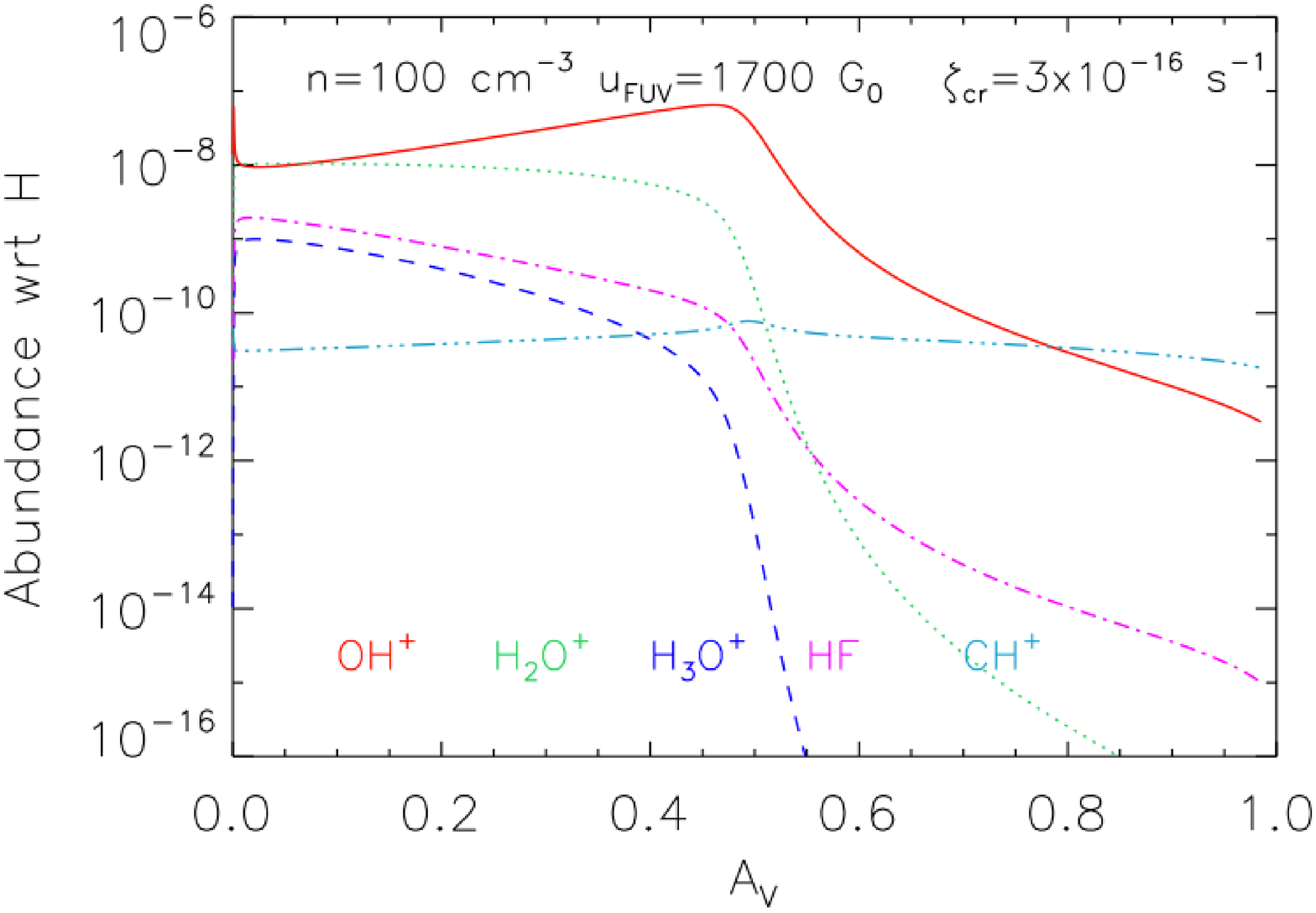} 
\caption{Abundance of OH$^{+}$, H$_2$O$^{+}$, H$_3$O$^{+}$, CH$^+$, and HF as a function of visual extinction, $A_\mathrm{V}$, predicted by the Meudon PDR code for a slab of gas with $n_\mathrm{H} = 100$~cm$^{-3}$, $\zeta_\mathrm{cr} = 3 \times 10^{-16}$~s$^{-1}$ and irradiated from the right by an FUV field of 1700~G$_0$.}
\label{fabu}
\end{center}
\end{figure}

We illustrate here the effect of a strong external FUV field such as the one implied by our modelling and observations, on the chemistry and the resulting molecular abundances of the species studied in this work. Fig.~\ref{fabu} shows these as a function of $A_\mathrm{V}$, as predicted by the Meudon PDR code for our best-fit model. The slab is exposed to an external FUV field of 1700~G$_0$ on one side (right), and to a typical FUV field of 1.7~G$_0$ on the other side (left). The predicted abundances are discussed in the following sub-sections, separated into families of species.

\subsubsection{Oxygen-bearing species: OH$^{+}$, H$_2$O$^{+}$, and H$_3$O$^{+}$}\label{O}

The strong incident FUV field affects the three O-bearing species in different ways: while it enhances the abundance of OH$^{+}$ at a certain depth, the amount of H$_2$O$^{+}$, and especially that of H$_3$O$^{+}$, are considerably reduced.

The formation of these molecules is believed to begin with the production of H$^{+}$ and H$_3$$^{+}$ via cosmic ray or X-ray ionisation of hydrogen, followed by charge transfer to produce O$^{+}$. Rapid hydrogen abstraction reactions of O$^{+}$ with H$_2$ follow to yield OH$^{+}$, then H$_2$O$^{+}$, and end with the production of H$_3$O$^{+}$. In diffuse molecular clouds, which have high electron abundances, H$_3$O$^{+}$ is destroyed via dissociative
recombination to yield OH and H$_2$O.

This is our case: the heavy UV irradiation on the slab surface gives rise to a high abundance of protons that favour the production of OH$^{+}$, but at the same time it destroys molecular hydrogen, thus hindering the cascade of reactions with H$_2$ that leads to the formation of H$_2$O$^{+}$ and subsequently to H$_3$O$^{+}$. The strong FUV field is therefore responsible for the relative column densities ($N_\mathrm{OH^+} > N_\mathrm{H_2O^+} > N_\mathrm{H_3O^+}$) we measure in the foreground slab.

\subsubsection{Hydrogen fluoride: HF}\label{HF}

The abundance of HF mirrors that of molecular hydrogen, with an estimated HF/H$_2$ abundance ratio within an order of magnitude from $3.6 \times 10^{-8}$ (Neufeld et al. \cite{neu05}). Indeed, HF has a higher binding energy than that of H$_2$ and, as a result, fluorine atoms are unique in reacting exothermically with the dominant constituent of interstellar molecular clouds, H$_2$, producing hydrogen fluoride via the reaction

\begin{displaymath}\rm F + H_2 \rightarrow HF + H \end{displaymath}~.

HF is therefore predicted to be the main reservoir of gas-phase fluorine in diffuse clouds where the amount of molecular hydrogen exceeds that of atomic hydrogen. This is reflected in Fig.~\ref{fabu}, where HF is less abundant close to the heavily irradiated surface of the slab, where the incident FUV photons dissociate H$_2$, thus preventing the above reaction to take place and form HF.

\subsubsection{Methylidyne cation: CH$^{+}$}

Unlike the O-bearing species and HF, the abundance of CH$^+$ does not change significantly with increasing depth within the slab (Fig.~\ref{fabu}) even though previous works (e.g. St\"auber et al. \cite{stau04}) predicted the abundance of CH$^+$ to be strongly enhanced with increasing UV field.

Our best-fit model for the foreground cloud (Sect.~\ref{fit}) predicts an abundance of about $2 \times 10^{-11}$ with respect to the total amount of hydrogen. As mentioned in Sect.~\ref{fit}, this is lower by a factor $\sim$250 than the observed value, a remarkably high discrepancy. It is likely that, as in the case of HF, there are several kinematical components hidden in the CH$^+$(1--0) line, although given its level of saturation, these are impossible to separate.

Nevertheless, if we assume that the entire CH$^+$ absorption comes from the foreground slab and adopt $N_\mathrm{H} = 1.9 \times 10^{21}$~cm$^{-2}$, which corresponds to that of the best-fit model, the CH$^{+}$ measured abundance is as large as $1.8 \times 10^{-8}$. This value is of the order of that found by other authors in analogous studies (Bruderer et al.~\cite{bru10}, Falgarone et al.~\cite{fal10a}, \cite{fal10b}), demonstrating, as stated also for instance by Falgarone et al.~(\cite{fal10a}) and Sheffer et al.~(\cite{sheffer08}), that the CH$^{+}$ abundances observed in the diffuse interstellar medium are several orders of magnitude higher than the predictions of UV-driven steady-state models, like the one we have used here. Non-equilibrium chemistry models need to be employed to correctly predict the CH$^{+}$ column densities. For example, Federman et al.~(\cite{fede96}) and Sheffer et al.~(\cite{sheffer08}) reproduced the observed CH$^+$ abundances by including non-thermal, magneto-hydrodynamic motions in their models, while Godard et al.~(\cite{god09}) achieved similar results by including supersonic turbulence in their chemistry modelling of diffuse interstellar gas, and illustrated in their Fig.~8 how standard PDR models fail to match the observed values, underestimating them by orders of magnitude.

Finally, it is important to bear in mind that the failure of the Meudon code to predict CH$^+$ column densities has effects on other families of species. For O-bearing molecules, the formation of CH$^+$ leads to CO via the following reactions (Federman et al.~\cite{fede96}):

\begin{displaymath}\rm CH^{+} + O \rightarrow CO^{+} + H \end{displaymath}

\begin{displaymath}\rm CO^{+} + H \rightarrow CO + H^{+} \end{displaymath}~.

Subsequent charge exchange between H$^{+}$ and O will lead to an enhanced amount of O$^{+}$, thus affecting the oxygen chemistry described in Sect.~\ref{O}. The inability of standard PDR models to correctly reproduce CH$^+$ abundances is therefore an important caveat for using such models, since it has deeper implications on the chemistry of other species, in particular the oxygen chemistry through the above-mentioned channel.

\subsection{Geometry of the slab}\label{geo}

An approximate calculation allows us to check whether the high FUV field necessary to explain our observations (between 1200 and 1700~G$_0$) is consistent. We assume the nearby Trapezium OB association, at a projected distance of 2~pc, to be responsible for this heavy irradiation. The Trapezium FUV output is dominated by four young stars: $\theta$~Ori~C, with a spectral type between O4 and O6 (Mason et al. \cite{mason98}, Baldwin et al. \cite{bal91}), and three early B-type stars, $\theta$~Ori~A, B, and D. Table~\ref{ttrapezium} presents the effective temperature (from Panagia~\cite{pana73}) and luminosity integrated over the FUV range, namely between 912 and 2400~\AA, for each of the four stars.

\begin{table}[!h]
\caption{Properties of the dominant Trapezium stars and distance from the absorbing slab}
\begin{tabular}{lccc}
\hline
Star & Spectral type & $T_\mathrm{eff}$ & $L_\mathrm{FUV}$\\
 &  & (10$^3$~K) & (10$^3$~L$_\odot$)\\
 \hline
$\theta$~Ori~A & B0.5 & 26.2 & 13\\
$\theta$~Ori~B & B1 & 22.6 & 6.3\\
$\theta$~Ori~C & O6--O4 & 42 -- 50 & 230 -- 1900\\
$\theta$~Ori~D & B0.5 & 26.2 & 13\\
\hline
 & \multicolumn{3}{c}{Distance (pc)} \\
 \cline{2-4}
 & $u_\mathrm{FUV} = 1200$~G$_0$ & \multicolumn{2}{c}{$u_\mathrm{FUV} = 1700$~G$_0$}\\
\hline
$\theta$~Ori~C: O6 & 2.0 & \multicolumn{2}{c}{1.7}\\
$\theta$~Ori~C: O4 & 5.4 & \multicolumn{2}{c}{4.6}\\
\hline
\end{tabular}
\label{ttrapezium}
\end{table}

A few parsec north of OMC-2~FIR~4 lies the NGC~1977 H{\sc ii} region, where the earliest-type ionising star is a B1, which is very faint in FUV output relative to the Trapezium. We therefore consider only the contribution of the four main stars of the Trapezium. Depending on whether $\theta$~Ori~C is an O6 or an O4 star, this results in a distance ranging from 2.0 to 5.4~pc between the Trapezium and the slab, for an irradiation of 1200~G$_0$, and from 1.7 to 4.6~pc for 1700~G$_0$ (Table~\ref{ttrapezium}). This is a rough estimate because we neglected the unknown contribution of dust extinction, which would yield a smaller distance. Despite this caveat, our calculation suffices to conclude that the high external FUV irradiation implied by our modelling and observations is consistent with the projected distance of 2~pc between OMC-2~FIR~4 and the Trapezium.

According to our best-fit model, the slab has a visual extinction of 1~magnitude or, in other words, a total hydrogen column density of $1.9 \times 10^{21}$~cm$^{-2}$. Taking into account its low density ($n_\mathrm{H} = 100$~cm$^{-3}$), this implies a line-of-sight size of 6~pc.

\subsection{Nature of the slab}

\begin{figure*}[!htb]
\begin{center}
\begin{tabular}{c}
\includegraphics[scale=0.4]{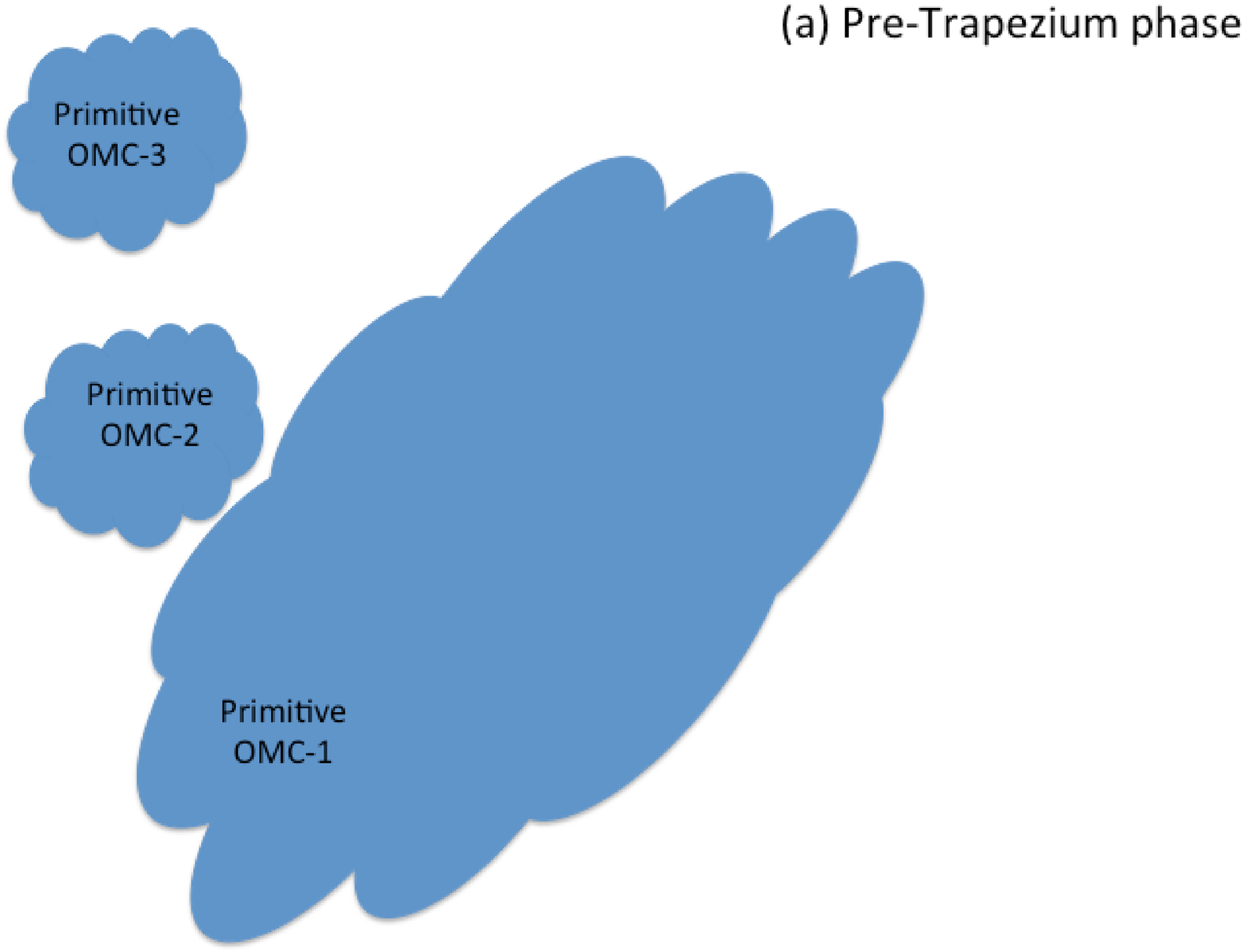}\\
\includegraphics[scale=0.45]{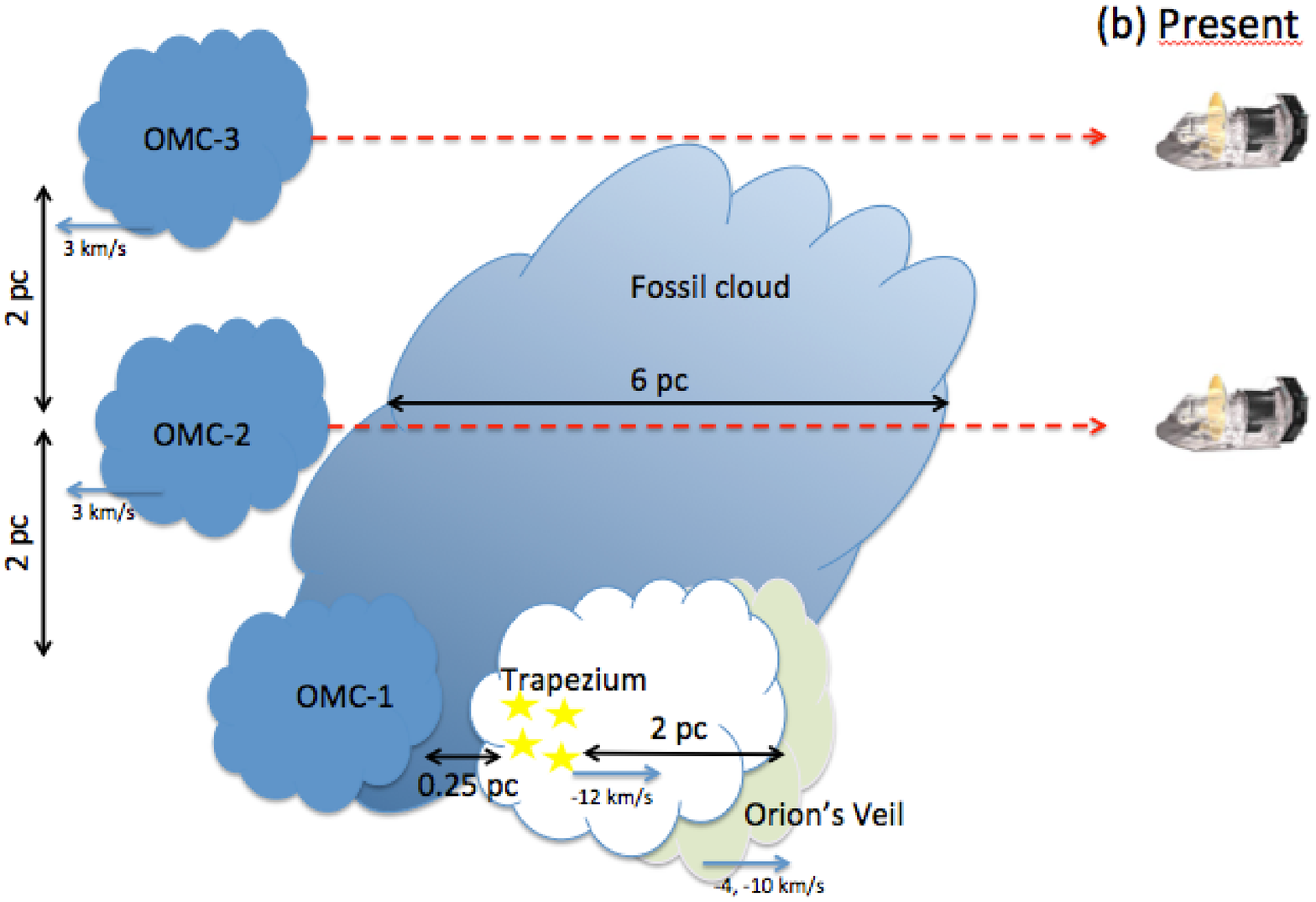}\\ 
\end{tabular}
\caption{Sketch of the OMC-1/2/3 region as it might have appeared before the formation of the Trapezium cluster (\textit{top}), and how it looks at present time (\textit{bottom}). The foreground cloud we detected is labelled ``fossil cloud", and the colour gradient across it represents a density gradient roughly in the north-south direction (see text). The sizes of each component and their velocities \textit{relative to the systemic velocity of OMC-1 and the foreground cloud} (i.e. $V_\mathrm{LSR} = 9$~km~s$^{-1}$) are marked. The observer, represented by a small picture of the \textit{Herschel} satellite, is situated to the right of each panel.}
\label{fgeo}
\end{center}
\end{figure*}

Numerous studies have been performed to investigate the foreground material towards OMC-1, which contains the Orion~KL star-forming region. These are mostly based on UV data of atomic and/or ionised metals, and HI absorption line observations at 21~cm. More recently, \textit{Herschel} observations have provided additional information (see e.g. Bergin et al. \cite{bergin10}). In contrast, no analogous research has been carried out towards OMC-2~FIR~4 and the rest of the OMC2/3 region, mostly because of the lack of strong UV background sources like the Trapezium in this direction. We can nevertheless compare our results with those found in past studies for OMC-1 in an attempt to search for links and understand the nature of the foreground cloud we observe. The foreground material towards the Trapezium and OMC-1 can be grouped into the following two components:

\begin{enumerate}
\item \textbf{Orion's Veil:} This term was used by O'Dell (\cite{odell01}) and Abel et al. (\cite{abel04}, \cite{abel06}) to denote a cloud of predominantly neutral gas that lies between the Trapezium cluster and us. It is also known as the ``foreground lid" (e.g. Wilson et al. \cite{wil97}). Orion's Veil has been observed and mapped in HI at 21~cm by several authors (e.g. van der Werf \& Goss \cite{vw90}, Wilson et al. \cite{wil97}), and neutral and ionised metals have also been detected in absorption at UV wavelengths (reviewed in O'Dell \cite{odell01} and Abel et al. \cite{abel06}). This system appears to have two main components, named A and B, at 1~km~s$^{-1}$ and 5~km~s$^{-1}$, respectively. Photoionisation models performed by Abel et al. (\cite{abel04}, \cite{abel06}) located the Veil at a distance of about 2~pc in front of the Trapezium, with a density of $\sim$10$^3$~cm$^{-3}$ and a thickness of 1~pc.
\item \textbf{C$^+$ interface between Orion KL and the Trapezium:} This component has been observed in carbon recombination lines tracing a C$^{+}$ region at 9~km~s$^{-1}$ (Balick et al. \cite{bal74}), and more recently in absorption at the same velocity in OH$^{+}$ and H$_2$O$^{+}$ (Gupta et al. \cite{gupta10}), and HF (Phillips et al. \cite{phil10}) with \textit{Herschel}-HIFI. The velocity is closer to the systemic velocity of the molecular gas in Orion KL ($V_\mathrm{lsr} = $8-9~km~s$^{-1}$, eg. Shimajiri et al. \cite{shim11}). This region is identified as a ``C$^{+}$ interface" between the Orion KL molecular gas and the Trapezium H{\sc ii} region. This component has strikingly similar characteristics to those seen towards OMC-2~FIR~4 in terms of detected species, velocity, and relative measured column densities.
\end{enumerate}

A summary of the properties of these foreground components, including OMC-2~FIR~4 from this work, is reported in Table~\ref{tveil}.

\begin{table*}[!tbh]
\begin{center}
\caption{Properties of the foreground material towards OMC-2~FIR~4 and towards Orion KL.}
\label{tveil}
\vspace{-3mm}
\begin{tabular}{lcccc}
\hline
 & OMC-2~FIR~4 & Orion KL\tablefootmark{a} & Orion's Veil (A)\tablefootmark{b} & Orion's Veil (B)\tablefootmark{b}\\
\hline
Species & OH$^+$, H$_2$O$^+$ & OH$^+$, H$_2$O$^+$ & HI, metals & HI, metals\\
 & HF, CH$^+$, C$^+$ & HF, C$^{+}$  & & \\
$V_\mathrm{LSR}$ (km~s$^{-1}$) & 9 & 9 & 1.3 & 5.3 \\
$n_\mathrm{H}$ (cm$^{-3}$) & $\sim$100& $\sim$1000 & 300 & 2500\\
N$_\mathrm{HI}$ (cm$^{-2}$) & $\sim$10$^{21}$ & $\sim$10$^{21}$ & $1.6 \times 10^{21}$ & $3.2 \times 10^{21}$\\
Thickness (pc) & 6 & $0.5$ & 1.3 & 0.5\\
\hline
\end{tabular}
\\
\tablefoottext{a}{Gupta et al. (\cite{gupta10}), Phillips et al. (\cite{phil10})}\\
\tablefoottext{b}{Abel et al. (\cite{abel04}, \cite{abel06}) and references therein}
\end{center}
\end{table*}

The complexity of the velocity field in Orion A has been noted by several authors (see review by O'Dell 2001). In a simplified picture, it reduces to the following three $V_\mathrm{LSR}$ components:
\begin{itemize}
\item Trapezium H{\sc ii} region: --3 km~s$^{-1}$
\item Foreground lid: 1 and 5 km~s$^{-1}$
\item dense molecular gas (Orion KL): 9~km~s$^{-1}$, with a south-north velocity gradient along the ISF (see Sect.~\ref{intro}).
\end{itemize}

Wilson et al. (\cite{wil97}) proposed a picture (see their Fig.~ 7) in which the Trapezium drives a strong wind, creating a cavity of low-density gas around it that moves towards us. This expansion is halted \textit{behind} the Trapezium by the high-density molecular gas, but is allowed to extend \textit{in front of it}, where there is no dense gas material to stop it. Beyond this cavity lies the ionised gas region, which is constrained between the foreground lid and the C$^{+}$ interface behind it. The foreground lid's receding velocity with respect to that of the H{\sc ii} region is interpreted as a flow away from this neutral material towards the lower pressure ionised gas in which the Trapezium is contained. The C$^{+}$ interface, also observed in absorption against the background dust continuum emission of Orion KL in OH$^{+}$, H$_2$O$^+$ and HF, as mentioned above, is confined between the Trapezium and the molecular gas, which are about 0.5~pc away from each other (Zuckerman \cite{zuc73}). The results found by Gupta et al. (\cite{gupta10}) suggest a density of about 10$^3$~cm$^{-3}$ for this interface, and possibly a high ionisation rate ($\zeta \sim 10^{-14}$~s$^{-1}$).

The similarity of our slab properties (velocity, column density) to those of the C$^{+}$ interface component lends support to a scenario in which our slab is an extension of that neutral gas region. The slab is blown-up to a lower density (100~cm$^{-3}$) and larger line-of-sight size (6~pc) by to the absence of elements that would confine it into a smaller and denser area, as for the C$^{+}$ interface, which is strongly affected by the Trapezium.

The absence of OH$^{+}$ absorption towards OMC-3, after comparison between the models and the derived upper limit on $N_\mathrm{OH^+}$ (Table~\ref{tobs}), can be interpreted by either of the following two possibilities: (i) the foreground cloud does not extend much farther to the north of OMC-2; (ii) it does reach the foreground of OMC-3 but is shallower ($A_\mathrm{V} \lesssim 0.5$) and/or more weakly irradiated ($u_\mathrm{FUV} \sim 1, ..., 200$~G$_0$) than the gas seen towards OMC-2. This is consistent with the larger distance of OMC-3 from the Trapezium cluster.

The cosmic ray ionisation rate implied by our observations ($\zeta = 3 \times 10^{-16}$~s$^{-1}$) is consistent with the typical value measured in diffuse clouds (Indriolo et al. \cite{indriolo}).

We stress that the foreground gas we detect towards OMC-2~FIR~4 has not been reported previously and therefore adds to the large ensemble of components present in the Orion~A complex. We interpret this foreground gas seen towards both OMC-1 and OMC-2 as the remnants of the OMC-1 parental molecular cloud, which once contained the primordial material that later allowed the formation of the Trapezium, Orion KL, and the BN infrared cluster. Figure~\ref{fgeo} presents two schematic cartoons of how the OMC-1/2/3 complex might have looked like before the Trapezium formed (\textit{top}) and how it currently appears according to our interpretation of the foreground slab, labelled ``fossil cloud" (\textit{bottom}).

\section{Summary}\label{conclusion}

As part of the \textit{Herschel} key programme CHESS, we have analysed a collection of absorption lines seen towards OMC-2~FIR~4. These include the ground-state transitions of OH$^{+}$, H$_2$O$^+$, HF, and CH$^{+}$. Our results can be summarised as follows:

\begin{enumerate}
\item The detected absorption lines peak at 9~km~s$^{-1}$, blue-shifted by 2~km~s$^{-1}$ with respect to the systemic velocity of OMC-2~FIR~4 ($V_\mathrm{LSR} = 11.4$~km~s$^{-1}$), indicating the presence of a foreground layer of gas in this direction.
\item Analysis using the Meudon PDR code shows that, regardless of whether we adopt standard interstellar medium extinction properties or those reported in the literature for the Orion Nebula, the slab is composed of predominantly neutral diffuse gas ($n_\mathrm{H} = 100$~cm$^{-3}$), which is heavily irradiated ($\sim 1500$~G$_0$) by an external source of FUV most likely arising from the nearby Trapezium OB association. The slab is 6~pc thick and bears many similarities with the so-called C$^{+}$ interface between Orion-KL, 2~pc south of OMC-2~FIR~4, and the Trapezium cluster.
\item Combining this work and previous studies carried out towards Orion~KL, we conclude that the foreground slab we discovered is the extension of the C$^{+}$ interface seen in the direction of Orion~KL, and interpret it to be the remains of the parental cloud of OMC-1, which extends from OMC-1 up to OMC-2.
\end{enumerate}

\begin{acknowledgements}
We are grateful to E. Roueff, J. Le Bourlot, F. Le Petit, and L. Wiesenfeld for their priceless help with the execution of the Meudon PDR code. A.L.S. and	ÊC.C.	 acknowledge funding from the CNES (Centre National d'\'Etudes Spatiales) and from the Agence Nationale pour la Recherche (ANR), France (project FORCOMS, contracts ANR-08-BLAN-022). M.K. gratefully acknowledges funding from an NWO grant, NOVA, Leids Kerkhoven-Bosscha Fonds and the COST Action on Astrochemistry. C.D. acknowledges funding from Leids Kerkhoven-Bosscha Fonds. A.F. has been partially supported within the programme CONSOLIDER INGENIO 2010, under grant CSD2009-00038 ``Molecular Astrophysics: The Herschel and ALMA Era ASTROMOL". 
\end{acknowledgements}

\end{document}